\documentclass[%
 reprint,
nofootinbib,
 amsmath,amssymb,
 aps,
prd,
]{revtex4-2}
\usepackage{amsmath}   
\usepackage{graphicx}
\usepackage{graphicx}
\usepackage{dcolumn}
\usepackage{bm}
\usepackage{hyperref}
\usepackage{aas_macros}
\usepackage{xcolor}
\usepackage{algorithm}
\usepackage{algpseudocode}
\usepackage{orcidlink}

\usepackage{graphicx}
\usepackage{tikz}
\usetikzlibrary{positioning}
\usepackage{caption}
\usepackage{graphicx}
\usepackage{subcaption}

\begin{document}


\title{Weak Lensing by Photometric Density Ridges}

\author{Mehraveh Nikjoo \orcidlink{0000-0001-8137-6699}\ \ }
\email{mehraveh.nikjoo@phdstud.ug.edu.pl}
\affiliation{Institute of Theoretical Physics and Astrophysics, University of Gdansk, 80-308 Gdansk, Poland}

\author{Joe Zuntz \orcidlink{0000-0001-9789-9646}\ \ }
\email{joe.zuntz@ed.ac.uk}
\affiliation{Institute for Astronomy, University of Edinburgh, Blackford Hill, Edinburgh EH9 3HJ, United Kingdom}

\author{Ben Moews \orcidlink{0000-0003-0897-040X}\ \ }
\email{ben.moews@ed.ac.uk}
\affiliation{Business School, University of Edinburgh, 29 Buccleuch Pl, Edinburgh EH8 9JS, United Kingdom}
\affiliation{Centre for Statistics, University of Edinburgh, Peter Guthrie Tait Rd, Edinburgh EH9 3FD, United Kingdom}

\begin{abstract}
Ridges in galaxy density fields measured by photometric surveys are 2D projections of filaments in the cosmic web, and so should lens light from background galaxies. We report on a detection of this effect in Dark Energy Survey Year 3 data at high significance, though not independently of galaxy-galaxy lensing. We describe improvements to the existing subspace-constrained mean shift algorithm to locate these ridges efficiently at scale, and examine the dependence of the signal in simulations on cosmological and algorithmic parameters. We find that it depends primarily on $S_8=\sigma_8 \left( \Omega_m / 0.3 \right)^{1/2}$, and discuss improvements to our methodology that would be needed to allow precision parameter estimation.

\end{abstract}

\maketitle

\section{\label{sec:intro}Introduction}

Predicted by General Relativity, \textit{gravitational lensing} occurs as a consequence of the bending of light as it travels through the curved spacetime created by massive objects. \textit{Weak lensing} manifests as a subtle distortion in the shape of background galaxies, which can be detected statistically across large populations of galaxies and encodes information on the integrated mass distribution and offers a unique window for probing large-scale structure and cosmology\citep{wl-review}. 

The primary set of statistics used to quantify weak lensing and extract this information are two-point correlation functions in either real or Fourier space. One statistic, \textit{cosmic shear}, uses galaxy shapes alone, correlating shapes among different redshift bins and measuring lensing mass structure correlations along the line of sight directly. Another, usually called galaxy-galaxy lensing, uses a second foreground sample of galaxies designed to trace high mass halos along the line of sight, and then measures the distortion by lensing around these foreground samples. Both, and their combination, have been measured in a series of current generation (\textit{Stage III}) surveys \citep{des-y3-bias,kids1000,hsc-3x2pt,lensing-without-borders} to great effect, and are key targets of upcoming (\textit{Stage IV}) surveys like the Rubin Observatory \citep{rubin-overview}, Euclid Space Telescope \citep{euclid-overview}, Roman Space Telescope \citep{roman-overview}, and the China Space Station Telescope \citep{csst-overview}.

Two-point measurements capture all the information present in purely Gaussian random fields, but non-linear evolution of cosmic structure adds additional higher order information. Additional statistics are needed to make use of this information, and in recent years, there has been a significant effort to transition towards higher-order and morphology-sensitive statistics to extract it. These include bispectra, three-point correlation functions, Minkowski functionals, peak counts, or moments of the convergence field have been developed to capture information from nonlinear structure formation \citep[see][for a review]{ajani2023}. These tools are sensitive to non-linear clustering, halo morphology, and void/filament topology, and can help break degeneracies between cosmological parameters. 

This work focuses on lensing by ridges, the photometric (two-dimensional) projection of filaments.  Lensing by 3D filaments found with spectroscopy has been studied before, with many works finding signals of lensing by individual filaments \citep{Epps,Higuchi,HyeongHan,Jauzac,Dutta}, or considering simulated data\citep{maturi,mead}. Several others have focused on searches for wide-field lensing signals around straight lines connecting Luminous Red Galaxy (LRG) samples in foreground spectroscopic samples, using estimators that remove the lensing from the LRGs themselves. This essentially treats all filaments as straight lines. \citet{clampitt} did so using Sloan Digital Sky Survey (SDSS) data for both the foreground and background samples, finding a $4.5\sigma$ detection of filament lensing, and comparing their fit to different models of filament thickness. \citet{kondo} used SDSS and Baryon Oscillation Spectroscopic Survey (BOSS) foreground samples and HyperSuprimeCam (HSC), and found similar results at $3.9\sigma$. \citet{xia} performed a similar analysis using a compilation of Kilo-Degree Survey, Red Cluster Sequence Lensing Survey, and the Canada-France-Hawaii Telescope Lensing Survey data, making a $3.4\sigma$ detection.

In this work we will consider a new statistical metric: lensing by photometrically-located filamentary structures in the foreground. This variant of galaxy-galaxy lensing has not previously been explored. An advantage of this method is much higher signal-to-noise, since we have a much higher density of foreground points. But since the nulling approach used in previous works relies on assuming that filaments are straight lines, our approach has the disadvantage of not removing any signal associated with halos at the ends of the filaments, and is therefore not independent of traditional galaxy-galaxy lensing signals.

Recent work in \citet{euclid-ridges} has compared ridge fields extracted from projected photometric maps to the underlying three-dimensional filament network by extracting skeletons from simulated galaxies' true redshifts and their photometric redshift, then comparing the two to quantify their similarity. They find that ridges are good tracers of 3D filaments, suggesting that this type of lensing should contain real new cosmological information rather than simply reflecting chance alignments of halo structures.

Prior work on density ridges in a cosmological context also includes the application to dark matter particles in hydrodynamical simulations and, subsequently, galaxy photometry of the Sloan Digital Sky Survey (SDSS) \citep{Chen_2015, Chen_2016}. The topic of lensing imprints by 3D filaments has also previously been investigated by \citet{He_2017} for the cosmic microwave background, using the separate datasets provided by \citet{Chen_2016} and the Planck CMB lensing potential map \citep{Planck_2014}. 

Various algorithms can be used to locate ridges in 2D or beyond. Those applied in cosmology include DISPERSE \citep{disperse} and machine learning approaches like MaLeFiSenta \citep{MaLeFiSenta}. In this work we use the subspace-constrained mean shift (SCMS) algorithm, as implemented in the DREDGE package introduced in \citet{Moews_2020} which extended the algorithm to include spherical distance metrics, mesh size and bandwidth optimisation, and an open-source implementation of the SCMS algorithm as a software package. It was applied in  \citet{Moews_2021} to locate ridges in lensing convergence maps, and here we apply it to find ridges in foreground lens maps, and then search for lensing signals around the ridges it finds. In general, the SCMS algorithm works in any dimensionality, and was applied to 3D filament finding in SDSS in \cite{carron_2021}.


The structure of this work is as follows.  In Section \ref{sec:methodology}, we describe our methodology for defining ridges from a foreground galaxy sample, and then computing the lensing signal with respect to them. In Section \ref{sec:data}, we describe the simulated and real data sets to which we apply the algorithm. In Section \ref{sec:results1} we present the results on noiseless simulations testing effects of cosmological and algorithmic parameters, and in Section \ref{sec:results2} we show measurements on real data and equivalent noisy simulations. We conclude in Section \ref{sec:conclusion}.

\section{\label{sec:methodology}Methodology}

\begin{figure}[h!]
    \centering
    \includegraphics[width=0.45\textwidth]{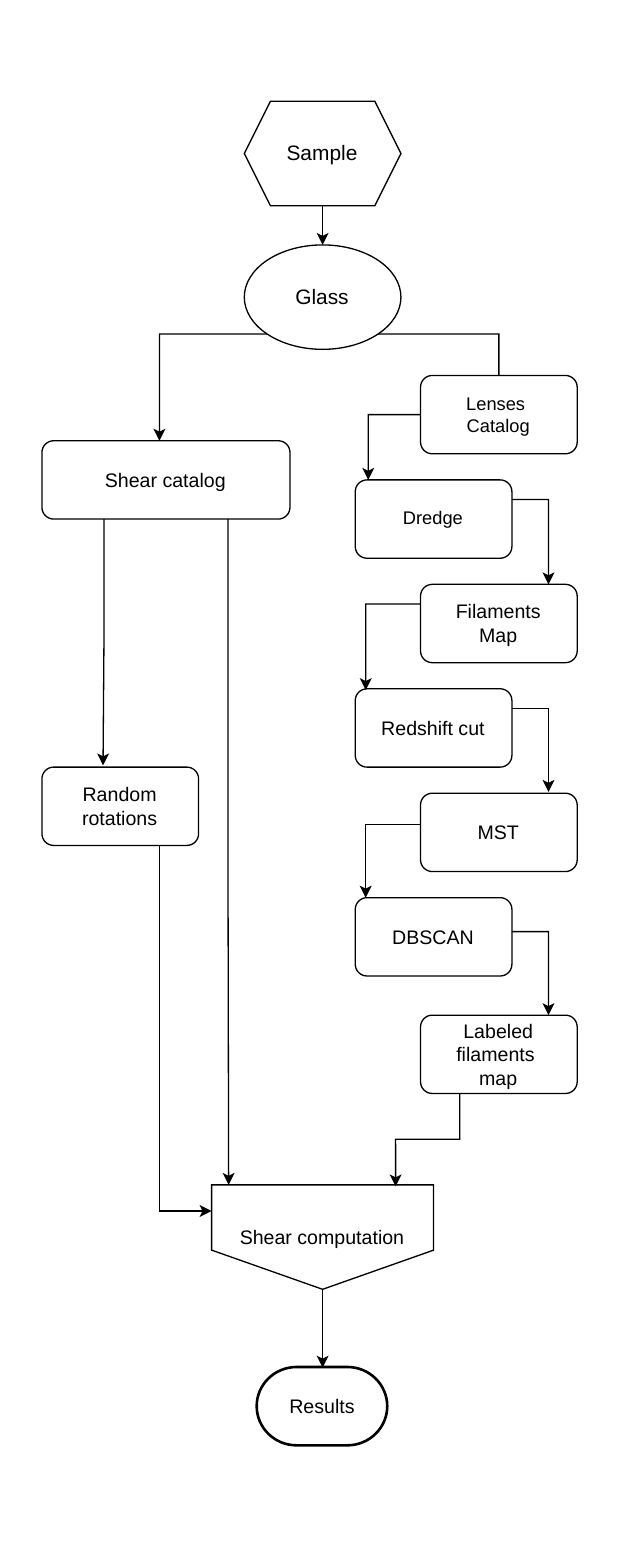}
    \caption{A schematic of the simulation and analysis procedure used in this paper.}
    \label{fig:flowchart}
\end{figure}

Figure \ref{fig:flowchart} presents a schematic overview of the steps in our methodology, from simulating a set of initial catalogues to computing the shear. In this section we detail the various stages.

\subsection{The SCMS Algorithm}

\subsubsection{Definition of a ridge}
\label{sec:ridge}

One-dimensional ridges in a D-dimensional space are defined mathematically as the set of points where (i) the Hessian matrix of the surface has $D - 1$ negative eigenvalues, and (ii) the projection of the surface gradient onto the associated $D - 1$ eigenvectors is zero \citep{Genovese_2014}. Points satisfying these conditions match the everyday definition of a extended narrow crest on a surface.

Ridges can be visually recognised quite straightforwardly, but robustly and efficiently locating them from discrete samples and/or noisy data can prove challenging. In this work, we use the \texttt{subspace-constrained mean shift} (SCMS) algorithm, an iterative approach to ridge estimation from point samples introduced in \citet{Ozertem_2011} and refined in \citet{Genovese_2014}. A more technical definition of ridges in this context for the interested reader can be found in Appendix~\ref{app:ridge}.

\begin{table*}[htbp]
\centering
\caption{Fiducial parameters used throughout this work. Algorithmic parameters control the density–ridge reconstruction, while cosmological parameters specify the fiducial simulation.}
\label{tab:fiducial_params}
\begin{tabular}{l c l}
\hline\hline
Parameter & Fiducial value & Description \\
\hline
\multicolumn{3}{c}{\textit{Ridge reconstruction parameters}} \\
\hline
Bandwidth / arcmin & $6$ & Kernel smoothing scale \\
Mesh size & $10^6$ & Number of initial mesh points \\
Density threshold & $15$ &  Min ridge density percentile to retain \\
$n_{\mathrm{neighbors}}$ & $5000$ & Nearest neighbors per update \\
Convergence / arcmin & $0.034$ & Iteration convergence threshold \\
Max unconverged fraction & $0.005$ & Stopping criterion \\
Max iterations & $5000$ & Maximum number of iterations \\
Distance metric & Haversine & Angular distance definition \\
\hline
\multicolumn{3}{c}{\textit{Cosmological parameters (fiducial simulation)}} \\
\hline
$\Omega_m$ & $0.3$ & Total matter density \\
$\Omega_b$ & $0.045$ & Baryon density \\
$H$ & $70$ km/s/Mpc & Hubble parameter \\
$\sigma_8$ & $0.8$ & Amplitude of matter density fluctuations \\
\hline\hline
\end{tabular}
\end{table*}

\subsubsection{SCMS algorithm details}
\label{sec:scms_algorithm}

The SCMS algorithm makes use of a kernel density estimate (KDE) of the distribution in the data space, using a radial basis function (RBF) kernel to identify points matching the definition in Subsection \ref{sec:ridge} in the resulting density profile.

Points subsequently forming the ridges start out as a grid or cloud of mesh points distinct from the data samples, for equidistant or uniformly-random distributed points, respectively. As noted by \citet{Moews_2020}, there is an analogy to be made between the operation of the algorithm and large-scale structure formation by gravitational forces. Sample points in the space evolve over time, or in this case through iterations of the algorithm. Starting with this mesh of potential ridge points overlaid on the data space, they are iteratively pulled towards high-density areas in a way that directs them towards ridges rather than purely peaks, forming a ``large-scale structure'' of the data.

The one-dimensional nature of these curvilinear structures is particularly useful for applications requiring an exact location rather than a constrained but spread-out profile, as is the case with variations of wavelets as an alternative \citep[see, for example,][]{hergt2017, vafaeisard2017}. A pseudocode listing of the SCMS algorithm is shown in Appendix \ref{app:scms}. For an overview of the inner workings of the algorithm, see the pioneering work by \citet{Ozertem_2011} and the exploration of nonparametric ridge estimation by \citet{Genovese_2014}.

\subsubsection{Changes to the algorithm}
\label{sec:scms_changes}

Our implementation makes use of the package \textsc{Dredge}, as described in \citet{Moews_2020} and \citet{Moews_2021}. We have updated it in several ways to speed it up when applied to large data sets ($> 10^6$ points):
\begin{itemize}
    \item Since the SCMS algorithm is embarrassingly parallel with respect to the mesh points that gravitate towards the ridges, we use MPI \citep{MPI} to parallelize over them.
    \item For the same reason, we updated the convergence criterion for the system to apply to the movement of individual mesh points instead of to the average movement of the whole ensemble. This allows us to stop updating points once they have individually converged, speeding up the iterations towards the end of the algorithm.
    \item We use a Ball Tree -- a nearest neighbour search method -- to query galaxies near to mesh points to use for the update step, avoiding using very distant objects very far from the mesh that will contribute negligibly to the update.
    \item We added checkpointing and just-in-time compilation with \textsc{Numba} \citep{numba} to the implementation.
\end{itemize}

\subsection{Convergence criteria}

Our algorithm employs two complementary notions of convergence. One operates at the level of individual mesh points to control their evolution under SCMS flow and defines when a single point has effectively reached a stationary configuration, while the other determines when the iterative procedure as a whole is terminated. 

In the former case, after each iteration, the algorithm evaluates the total displacement applied to the point in coordinate space. If this displacement falls below a fixed tolerance, the point is deemed to have converged and is removed from further updates. In practice, a small subset of mesh points may continue to evolve very slowly or exhibit oscillatory behavior, requiring a global convergence at the level of all mesh points helps avoid excessive runtime. 

This global convergence is defined in terms of the fraction of mesh points that remain active. Once the proportion of unconverged points falls below a prescribed threshold the algorithm terminates.

\subsection{Density Thresholds}
As in \citet{Chen_2015}, we can apply a local number density cut to the ridges that we detect, to remove both spurious noise detections and genuine ridges which nonetheless have a low density and so a low lensing signal.

We do so by building a KDE of the full foreground galaxy sample, and using it to evaluate the density at the detected ridge points. We can then select a certain density percentile to retain. We explore the effects on our signal of choosing this percentile in Section \ref{sec:results:threshold}.

\subsection{Segmentation: Ridge Points to Filaments}
We wish to partition the ridge points found by the SCMS process into discrete groups representing individual ridges.
This is an ambiguous process in our case: there are multiple viable ways to split ridges, though most reasonable variants are expected to generate a similar lensing signal.

We do this by:
\begin{enumerate}
  \item Constructing a Minimum Spanning Tree (MST) from the ridge point set using nearest-neighbour distances. This joins all the ridge points together with connecting edges.
  \item Identifying branch points in the MST as nodes with degree $> 2$, and splitting the tree there. This splits the ridges at nodes where they connect.
  \item Applying the DBSCAN clustering algorithm within each MST segment to split into separate ridges and assign labels to ridge points.
\end{enumerate}

We begin with a discrete set of ridge coordinates tracing regions of enhanced projected density, with the goal of extracting geometrically coherent filamentary structures. In this framework, the Minimum Spanning Tree (MST) encodes the global topological connectivity of the point distribution. The MST connects all points while minimizing the total edge length and avoiding closed loops, thereby providing a unique, loop-free\footnote{While being loop-free is not necessarily a requirement of ridge structure, especially in 2D, underlying 3D filaments are expected to be largely so.} representation of the underlying filamentary skeleton.

Each point first identifies its k-nearest neighbors using a KDTree search, generating a sparse weighted graph whose edge weights correspond to the pairwise spatial separations of connected neighbors. Applying the MST algorithm to this graph yields a network that minimizes the cumulative edge distances while maintaining global connectivity.

Real cosmic filaments exhibit junctions and bifurcations that give rise to nodes connecting multiple subfilaments. To isolate single, physically coherent branches and reduce projection-induced blending in lensing analyses, we identify branch points as nodes with degree greater than two and remove them from the MST. The resulting disjoint connected components represent individual filament segments. From a weak lensing perspective, treating all branches as a single filament would increase the projected overlap and artificially mix the shear signal from multiple directions; the segmentation step therefore ensures that each recovered structure contributes a distinct, interpretable lensing signature.

The subsequent step performs density-based segmentation on each MST component using the DBSCAN (Density-Based Spatial Clustering of Applications with Noise) algorithm. DBSCAN groups points according to local spatial density without imposing assumptions about cluster shape or topology \cite{SCADDA}. Within each MST segment, this procedure consolidates locally connected ridges into continuous and physically coherent filamentary chains, while excluding spurious or sparsely sampled points classified as noise.

\begin{figure}[H]
\centering

\begin{tikzpicture}[
    node distance=1.4cm,
    every node/.style={align=center}
]

\node (ridge) {
    \includegraphics[width=0.9\columnwidth]{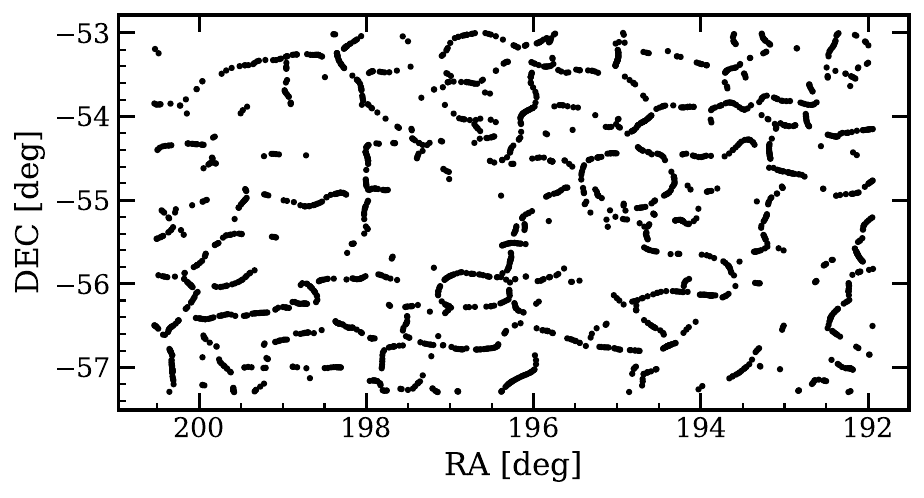}
};
\node[below=1mm of ridge] {\textbf{Ridge Points}};

\node[below=1.0cm of ridge] (mst) {
    \includegraphics[width=0.9\columnwidth]{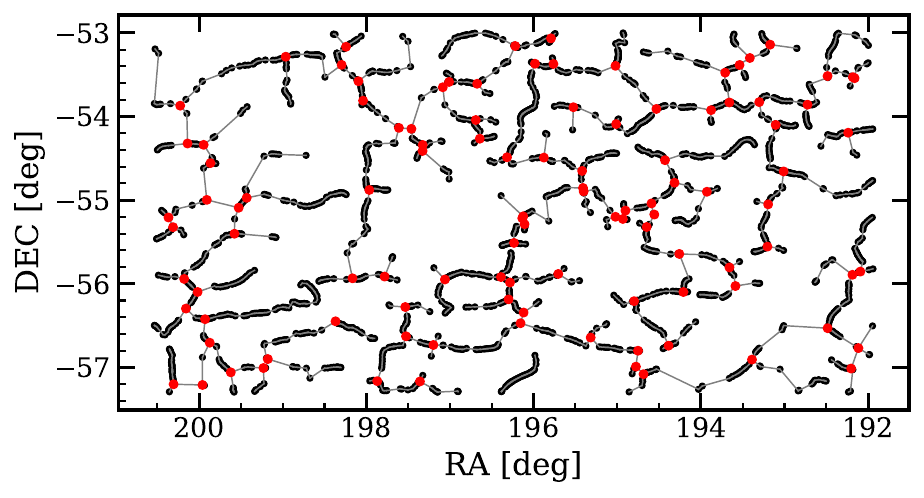}
};
\node[below=1mm of mst] {\textbf{MST + Branch Construction}};

\node[below=1.0cm of mst] (dbscan) {
    \includegraphics[width=0.9\columnwidth]{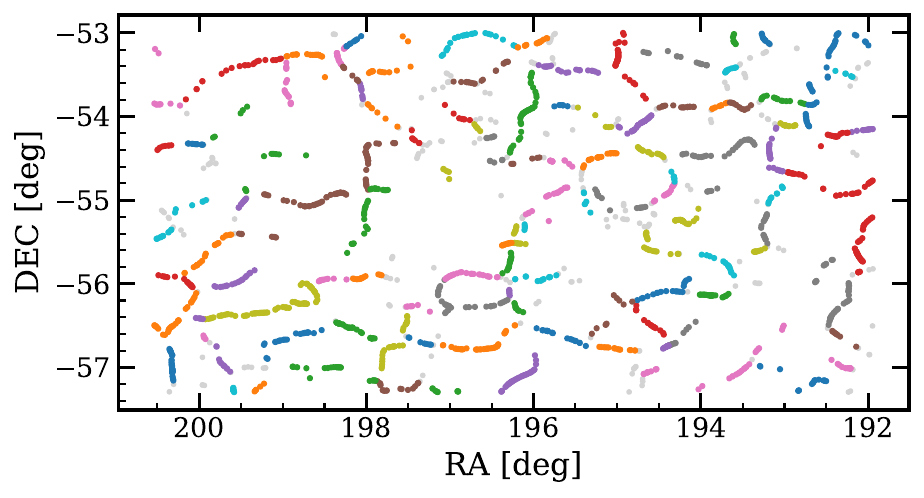}
};
\node[below=1mm of dbscan] {\textbf{DBSCAN Filaments}};


\end{tikzpicture}

\caption{
Schematic overview of the ridge-to-filament identification pipeline in Radians.
}
\label{fig:pipeline}

\end{figure}
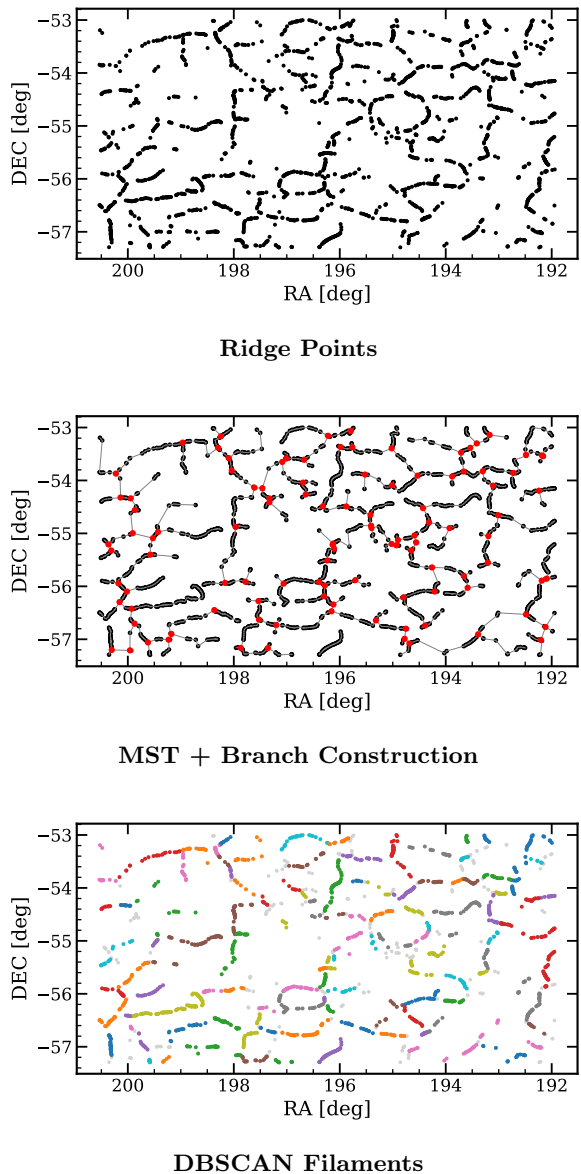

Figure \ref{fig:pipeline} shows and overview of these stages of ridge segmentation and labelling. 

\subsection{Tangential Shear measurement}

\begin{figure}[H]
    \centering
\centering
\includegraphics[width=0.8\linewidth]{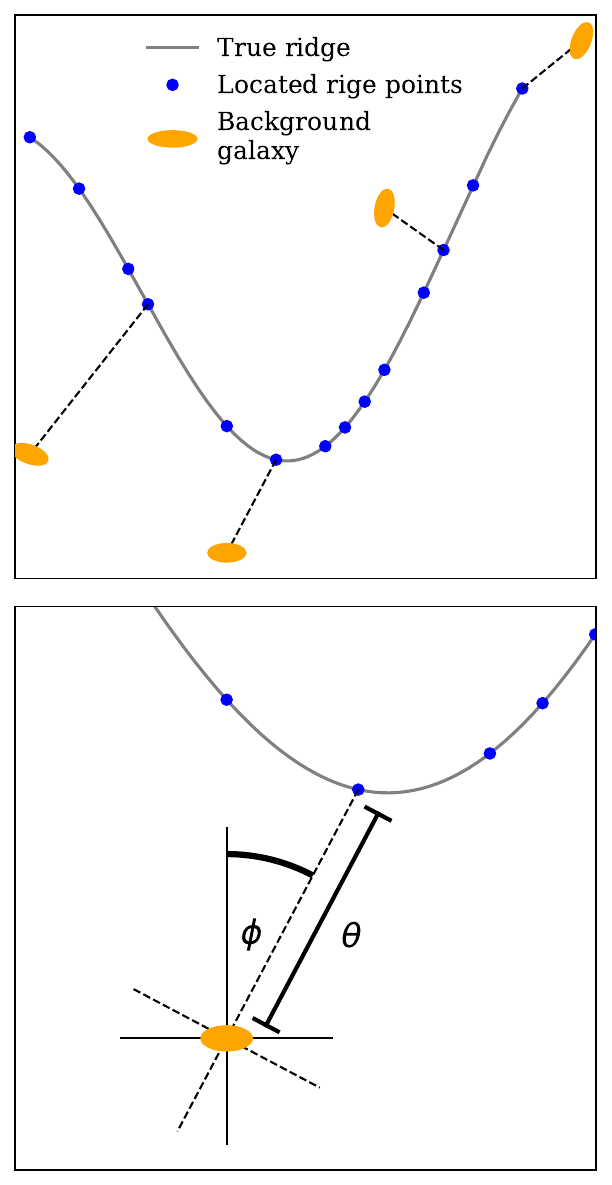}
\caption{Top: a schematic of the identification of nearest points on a ridge to background galaxies. Bottom: A zoom-in on one pair showing the separation angle $\theta$ and the rotation angle $\phi$ to convert to tangential shears. \label{fig:diagram}}
\end{figure}

Having identified the filamentary structure in the previous steps, we aim to quantify the lensing signal around each labelled filament.
Thus, we measure the tangential shear of background galaxies relative to each filament segment. A schematic of the geometry we consider is shown in Figure \ref{fig:diagram}.

The formalism follows standard treatments \citep{prat_bacon} where the observed ellipticity components ($\gamma_1, \gamma_2$) of source/background galaxies are related to the projected shear field through local coordinate rotations:

\begin{align}
    \gamma_+ &= -(\gamma_1 \cos{2\phi} + \gamma_2 \sin{2\phi}), \\
    \gamma_{\times} &= \gamma_1 \sin{2\phi} - \gamma_2 \cos{2\phi},
\end{align}

Where $\phi $ is the position angle between the source and the nearest point on the filament, which we compute using  \textsc{Astropy} \citep{astropy:2013,astropy:2018,astropy:2022}. Generally, in the absence of systematics, and assuming no parity-violating signal, $\langle \gamma_{\times} \rangle = 0$ is expected for standard galaxy-galaxy lensing. For ridge lensing we shall see the same, although this is not necessarily required in the case of non-straight ridges.

For each galaxy-filament pair, we identify the position angle between the source and the vector to the nearest point located on the filament. Pairs are then binned by the separation angle $\theta$ and the mean $\gamma_+$ and $\gamma_\times$ across all pairs is computed. To perform this calculation for a large map and reduce computational cost, the source catalog is first divided into low-resolution HEALPix pixels to avoid unnecessary whole-sky searches. All background galaxies projected into the filament neighbourhood are then selected using a Haversine metric, which defines the local sky used for shear calculation. 

We use 20 bins, logarithmically spaced from 1 to 60 arcmin, though at the smaller scales our log-normal simulations are not expected to reproduce realistic structure.

\section{\label{sec:data}Data \& Simulations}
\subsection{GLASS Simulations}

We use the package \textsc{GLASS} \citep{tessore23} to generate simulated realizations of our lens and source galaxy samples. We use it to generate log-normal density fields consistent with matter power spectra computed by \textsc{CAMB} \citep{camb1, camb2} up to $\ell_\textrm{max} = 10,000$ and $z_\mathrm{max}=3$ in shells of spacing 150 Mpc/h. It then computes the convergence in those shells, and generates source and lens samples consistent with both density, shear, and a selected shape noise level.

Our fiducial cosmological parameters are $\Omega_m = 0.3$, $\mathrm{H}_0=70$ km/s/Mpc, $\Omega_b=0.045$, and $\sigma_8=0.8$.  

\subsubsection{DES-like}
\label{sec:data:deslike}
\begin{figure}[h!]
    \centering
    \includegraphics[width=0.45\textwidth]{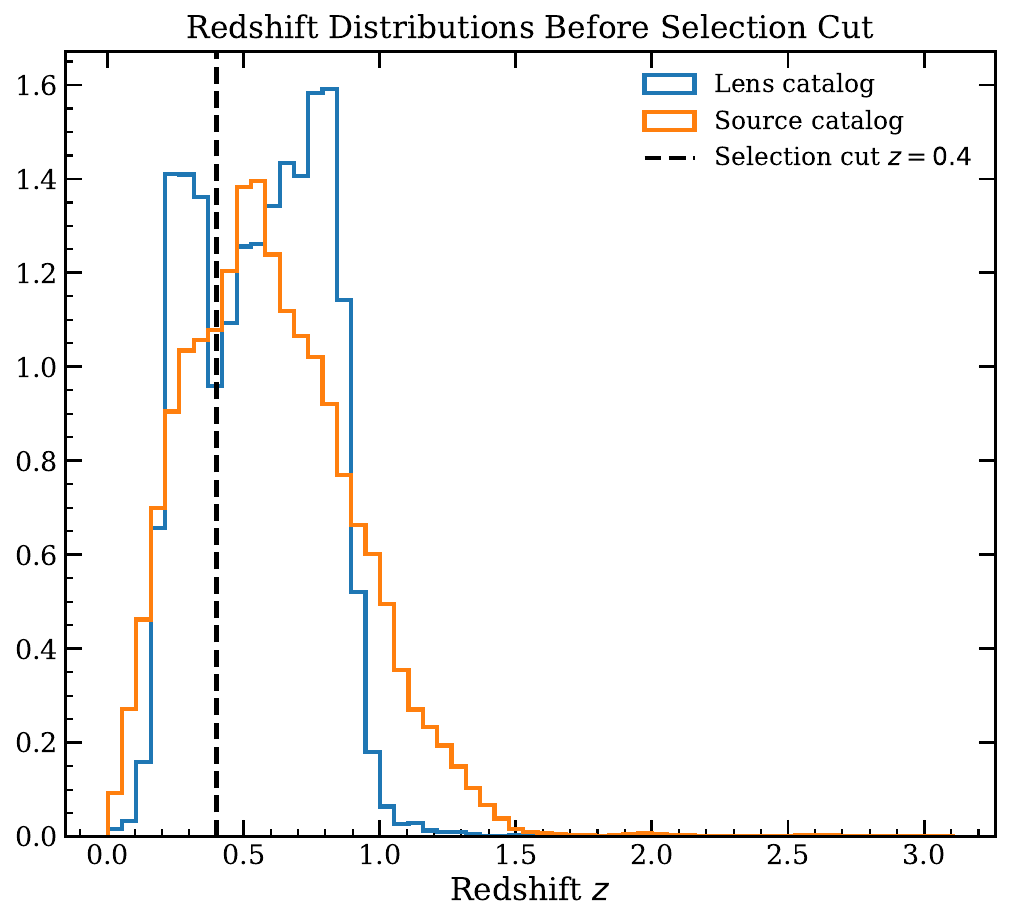}
    \caption{The input redshift distributions for the simulation foreground lenses (on whose density the ridges are defined) and background sources (whose orientation around the ridges we measure). The vertical line shows $z=0.4$: we cut the lenses to $z<0.4$ and the sources to $z>0.4$, defined on the true redshifts in the simulations, to maximize the signal-to-noise by ensuring that the sources are behind the lenses.}
    \label{fig:redshifts}
\end{figure}

For our primary DES-like simulations we use the Year 3 GOLD mask\footnote{\protect{https://des.ncsa.illinois.edu/releases/y3a2/Y3key-catalogs}} which is passed to \textsc{GLASS}. We supply the main non-tomographic redshift distributions of both the \textsc{Metacal} source sample \citep{des-y3-source-nz} and \textsc{MagLim} lens sample \citep{des-y3-lens-nz} as generated by the survey, as shown in Figure \ref{fig:redshifts}, and also match the object density of the samples. We use linear galaxy biases, with value given by the mean of those found in \citet{des-y3-bias}. We use a constant shape noise per component in the source sample, $\sigma_e=0.26$, though we also make signal-only simulations in which this set to $\sigma_e=10^{-3}$.

\subsection{DES Data}
\label{sec:data:des}
For our real DES data analysis we use the \textsc{MetaCal} \citep{metacal} catalog released as part of the DES Year 3 results\footnote{\url{https://desdr-server.ncsa.illinois.edu/despublic/y3a2_files/y3kp_cats/}} \citep{des-y3-shear}. We select galaxies in the same way as the main DES cosmic shear analysis, by requiring unflagged galaxies with signal-to-noise ratio $S/N>10$, size to PSF size ratio $T/T_\mathrm{PSF}>0.5$, and weight $w>0$.

Based on these selections we apply the metacalibration procedure from \citet{metacal} on the catalogs, finding a calibration factor $R=0.68$.

\subsection{Redshift cuts}

Since light can only be lensed by structures along its path, configurations where the filament is behind the source galaxy will only add noise to the measurement. To optimize any possible signal-to-noise, we therefore split both the lens galaxies used to generate the ridges and the source galaxies. We take lens galaxies with redshifts $z<0.4$, and source galaxies with $z>0.4$. 

We note that in on our real DES data this cut is performed on photometric redshifts generated using the DNF method \cite{dnf} and released as part of the DES Y3 GOLD data set \cite{des-gold}. In our simulations, however, they are performed on the known true galaxy redshifts. This makes the simulation cuts considerably more precise than those on real data.

We also note that we do not calculate an additional selection bias for these cuts in the \textsc{MetaCal} framework, because the DNF redshift catalog is not available for sheared variants of the catalogs. Based on the sizes of other selection biases, we expect this to induce a percent-level bias in our results, and future precision comparisons of ridge lensing to data should account for this bias.

\section{\label{sec:results1}Results: Noiseless Simulations}

In this section we study simulations containing no shape noise in order to test the convergence of algorithm parameters and examine the behaviour of the underlying shear signal with respect to cosmology. In this limit, the shear signal is determined entirely by the projected mass distribution and by the geometry of the ridge reconstruction algorithm. The purpose of this section is therefore not to optimize signal-to-noise, but to establish the robustness and internal consistency of the measurement pipeline under controlled conditions.

\subsection{General Signal Structure}
\begin{figure}[h!]
    \centering
    \includegraphics[width=0.45\textwidth]{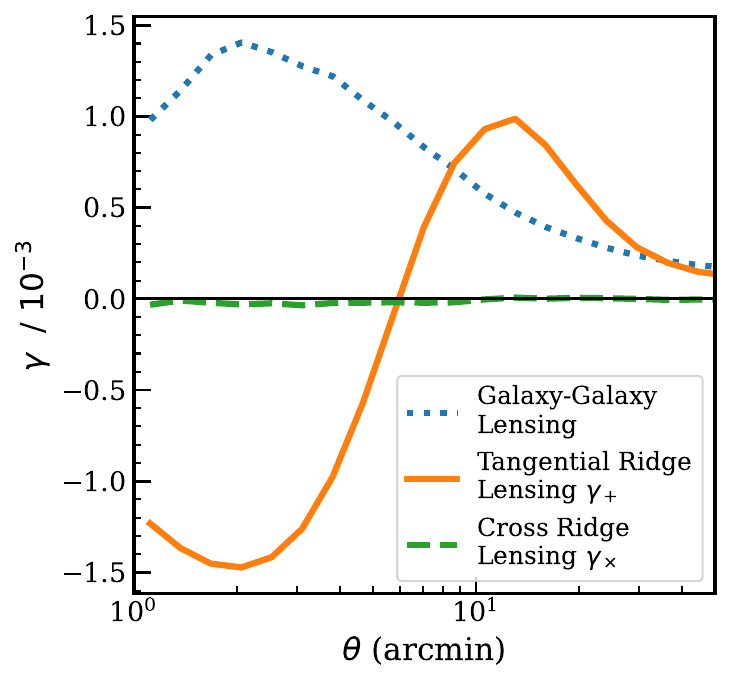}
    \caption{The tangential shear around ridges in a fiducial shape noise-free simulation as a function of separation, compared with the galaxy-galaxy lensing signal from the same data.}.
    \label{fig:ggl_comparison}
\end{figure}

Figure \ref{fig:ggl_comparison} shows the ridge lensing signal we measure on a noise-free simulation with our fiducial cosmological and algorithm parameters (discussed in subsequent sections). We compare it to a standard galaxy-galaxy lensing signal on the same objects. It can be seen that at large separations the signals approximately converge. At scales below the SCMS bandwidth (smoothing) parameter, set here to 6 arcminutes, the negative ridge lensing signal corresponds to background galaxies aligning perpendicularly along the ridge because they are primarily lensed by other structures along it, as seen in \citet{Higuchi} and \citet{clampitt}.

The cross-shear $\gamma_x$ is around two orders of magnitude smaller than $\gamma_t$ for ridges. Because this is fairly consistent throughout our results, and much smaller than the noise levels we shall see in section \ref{sec:results2}, we do not consider it in our parameter tests.

\subsection{SCMS Parameter Convergence}
We now test the convergence of our SCMS implementation with respect to the three main parameters that trade off numerical convergence against computational time (as opposed to the bandwidth and thresholding parameter, which change the physical meaning of the detected ridges).

\subsubsection{Target move distance}

The first parameter is a target move distance. If a mesh point moves less than the target from one iteration to the next then it is deemed to converged. We find that 98.9\% of points that have converged when the target is $10^{-5}$ degrees have also converged if the target is $10^{-4}$ degrees, so that any remaining change by decreasing this tolerance is likely to have negligible effect on the shear. There is no visually detectable difference between the ridges detected in the two scenarios. We adopt a target of $10^{-5}$ degrees in this paper.

\subsubsection{Number of neighbour galaxies}

\begin{figure}[h!]
    \centering
    \includegraphics[width=0.45\textwidth]{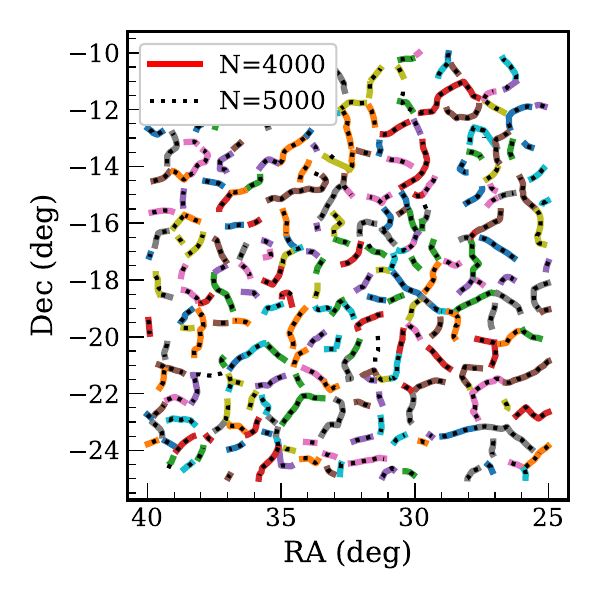}
    \caption{Detected ridges in a small region of our simulation when using 4000 and 5000 neighbours around each mesh point. There is good visual agreement between the two, indicating reasonable convergence.}.
    \label{fig:convergence_neighbours}
\end{figure}

Figure \ref{fig:convergence_neighbours} shows a check of the second parameter, the number of nearest neighbour galaxies to each ridge mesh point to use to update the ridge locations (the original algorithm used all the points in the entire simulation). While there are visible differences between $N=4000$ and $N=5000$, almost all of the detected ridges are unchanged between the two. We therefore adopt $N=5000$ in this paper.

\subsubsection{Effect of mesh size}
The mesh size parameter controls the initial number of uniformly distributed points from which the algorithm begins its iterative convergence process. The algorithm generates a random mesh of points spanning the spatial domain of the input galaxy distribution, which are then iteratively shifted towards the underlying density ridges. Each mesh point undergoes a successive update until the displacement between iterations falls below the convergence threshold. At this point, a ridge coordinate is considered as accepted. The mesh size is, thus, an indicator of the initial spatial sampling density with larger values providing more complete coverage at the cost of increased computational expense. 

\begin{figure}[t]
    \centering
        \centering
        \includegraphics[width=\linewidth]{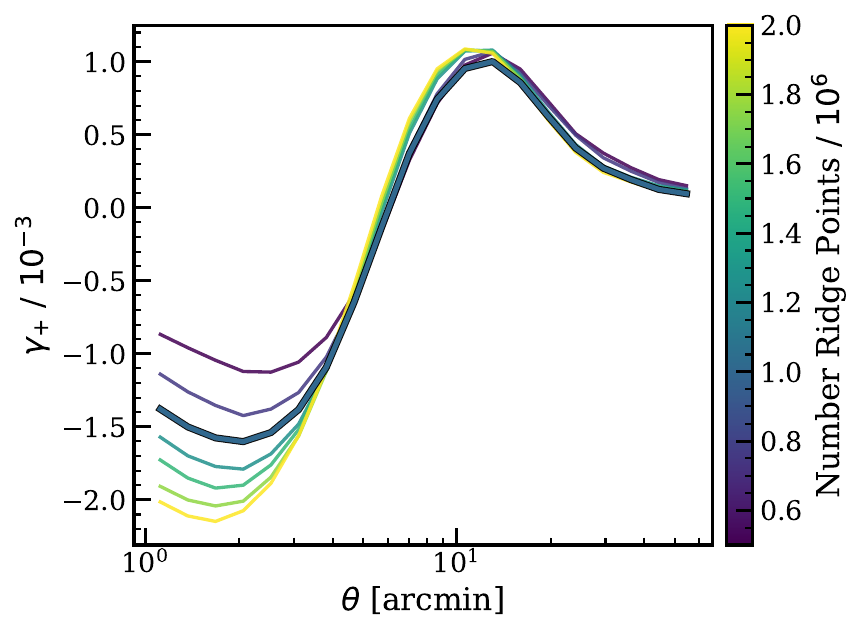}
    \caption{
        Variation of the signal-only tangential ridge shear signal with respect to the number of points used to locate ridges (the mesh size). The small-scale signal increases as more points map out ridge structure more clearly.
    }
    \label{fig:ShearVsMesh}
\end{figure}

To assess the sensitivity of our results to this parameter, we performed a systematic convergence study by running the complete pipeline for mesh threshold values of $N=5\times10^5$ to $2\times10^6$. For each value, we computed the tangential shear signal $\gamma_+$ around the constructed filaments. As shown in Figure \ref{fig:ShearVsMesh} the tangential shear profiles are consistent at large scales across all tested mesh sizes, with small variation at small angular scales. For this study, we adopt an intermediate mesh size $N=10^6$ to reduce the computational costs.

\subsection{Physical Parameters}
We now check the behaviour of the shape noise-free ridge lensing signal as a function of the two ridge detection parameters that determine the physical meaning of the ridges we find: the density threshold cut and the bandwidth.

\subsubsection{Effect of density threshold}
\label{sec:results:threshold}
We perform a density threshold cut on the detected ridge points, using a KDE to identify those that lie in low-density regions of the map. We parametrize this cut as a percentile, removing detected ridge points in regions below that level.

Physically, this selection is motivated by the expectation that higher-density ridges are both more robust against noise fluctuations and more efficient lenses. Increasing the density threshold enhances the lensing signal, but reduces the number of contributing structures. This leads to a trade-off between signal amplitude and statistical noise.

The tangential shear exhibits a clear dependence on the threshold. As the latter increases, the amplitude systematically rises, as expected, across a range of angular separations, while the overall shape of the profile remains stable. The location of the peak is preserved, and no qualitative change in the profile morphology is observed. Interestingly, this smaller scale negative signal remains mostly insensitive to density selections.

\begin{figure}[t]
    \centering
    \includegraphics[width=\linewidth]{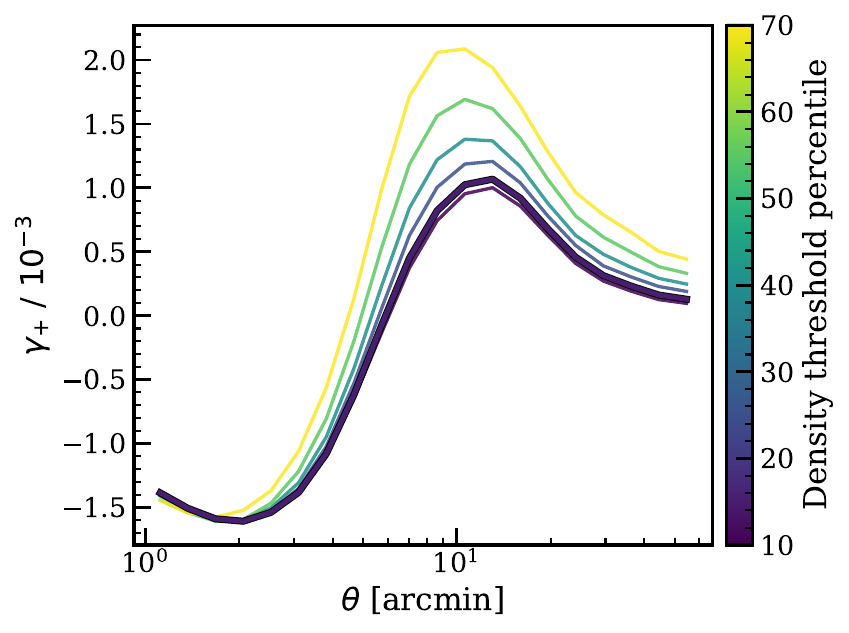}
    \caption{
    Variation of the signal-only tangential ridge shear signal with respect to a percentile used as density threshold; ridges with density below the threshold are excluded.
    }
    \label{fig:threshold}
\end{figure}

\subsubsection{Effects of bandwidth}
The bandwidth parameter serves as a smoothing scale in our algorithm. It governs how the galaxy distribution is mapped onto a density field. In the present analysis, the bandwidth parameter is used as the Gaussian kernel weighting and controls both the kernel density estimation and the subsequent subspace-constrained mean shift (SCMS) updates.  Therefore, bandwidth defines the scale at which filamentary or ridge-like structures are assumed to be physically meaningful.
As a primary step, an initial random mesh of candidate ridge points is also truncated by discarding points whose nearest data neighbour lies farther than a fixed multiple of the bandwidth. Primarily, the choice of bandwidth represent a trade-off between sensitivity to small-scale filamentary structure and the robustness against spurious noise detection.

The \textsc{Dredge} code has an internal estimator for an appropriate bandwidth from the cross validation maximum likelihood KDE bandwidth tool in the \textsc{statsmodels} package \citep{statsmodels}. We use that to obtain an initial order of magnitude for the bandwidth, and then vary around that value.

Once chosen, the bandwidth remains unchanged throughout the ridge-finding process; the SCMS iterations do not adapt or refine it.

\begin{figure}[t]
    \centering

    \includegraphics[width=\linewidth]{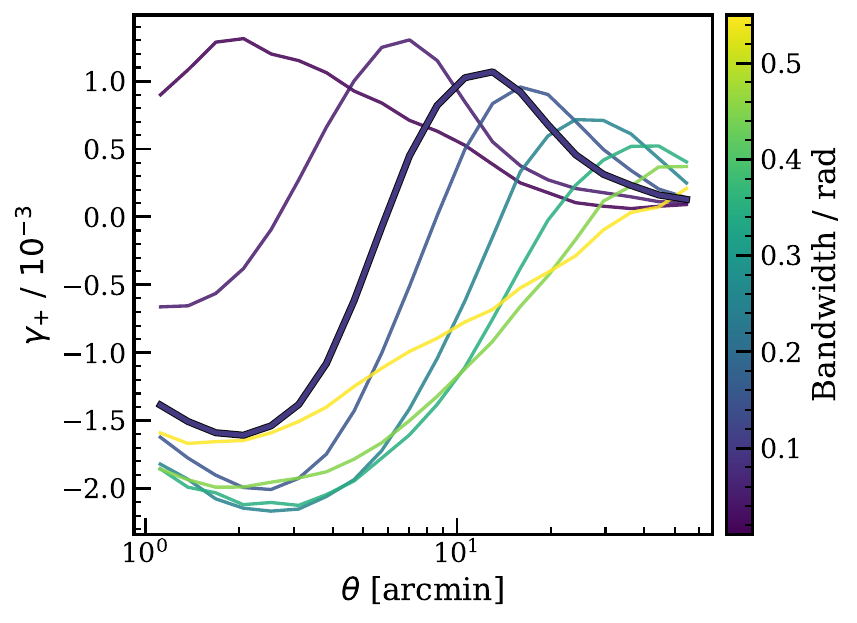}
    \caption{
Variation of the signal-only tangential ridge shear signal with respect to the bandwidth (smoothing) parameter. For small bandwidths the signal resembles the galaxy-galaxy lensing signal.
    }
    \label{fig:ShearVsBand}
\end{figure}

In Figure ~\ref{fig:ShearVsBand} Increasing the bandwidth suppresses the positive part of the signal, but has a more complicated behaviour in the small-scale negative part. The location of the peak also shifts towards larger angular separations; this shift is expected as larger bandwidths effectively produce more diffuse ridges and, thus, the angular scale at which the signal is maximized moves outward. The zero-crossing point is close to linearly dependent on the bandwidth ($\theta_0 \propto b^{1.04}$), as expected for a smoothing scale.

Finally, we note that the measurement obtained for the smallest bandwidth coincides with the galaxy-galaxy lensing signal shown in Figure ~\ref{fig:ggl_comparison}. This behavior is physically expected: as the bandwidth is reduced, the filament reconstruction becomes increasingly sensitive to overdensities in the galaxy field. In this limit, the dominant structures are very short filaments corresponding to  individual galaxies and their host halos.

\subsection{Cosmological signal}
We now explore how the signal we detect responds to the values of primary cosmological parameters, again in the noise-free case.

To isolate how the filament tangential shear responds to changes in cosmological parameters, we look at variations across the plane of $(\Omega_m, \sigma_8)$, both along the axes and along the line of structure growth parameter $S_8 = \sigma_8 (\Omega_m/0.3)^{1/2}$ and its orthogonal complement. 

We have fixed the previously mentioned parameters of the dredge code to the values in \ref{tab:fiducial_params}.

\begin{figure*}[t]
    \centering

    \begin{subfigure}{0.48\textwidth}
        \centering
        \includegraphics[width=\linewidth]{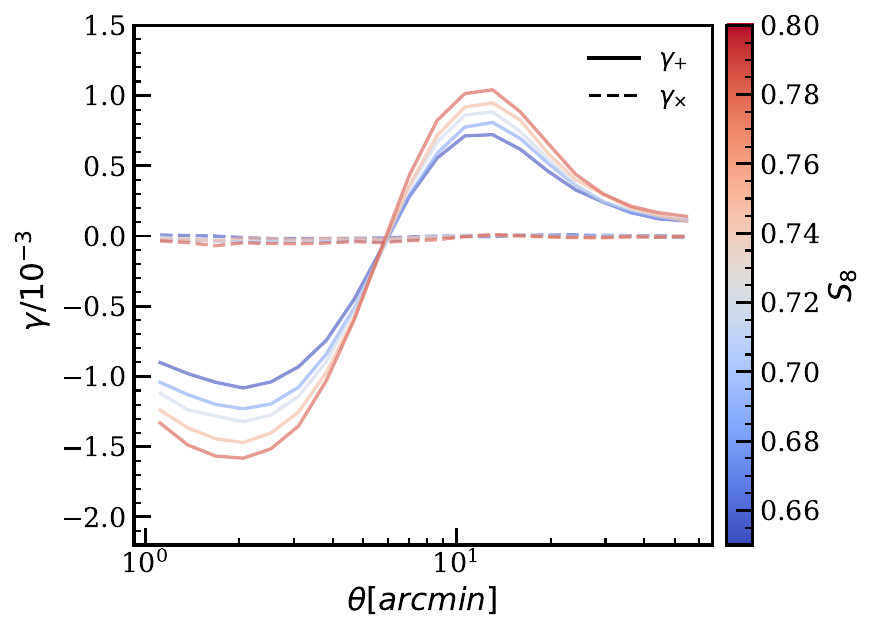}
    \end{subfigure}
    \hfill
    \begin{subfigure}{0.48\textwidth}
        \centering
        \includegraphics[width=\linewidth]{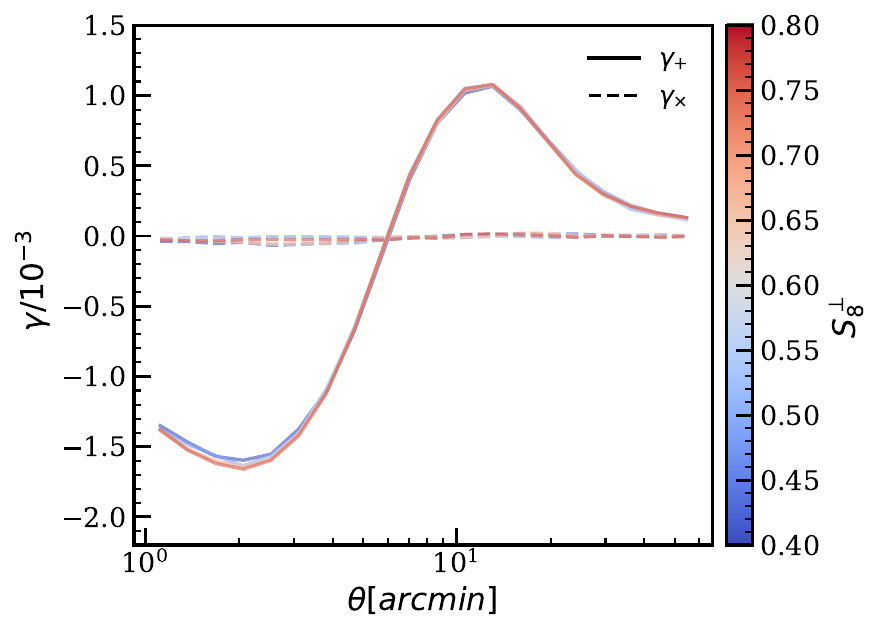}
    \end{subfigure}

    \vspace{0.6em}

    \begin{subfigure}{0.48\textwidth}
        \centering
        \includegraphics[width=\linewidth]{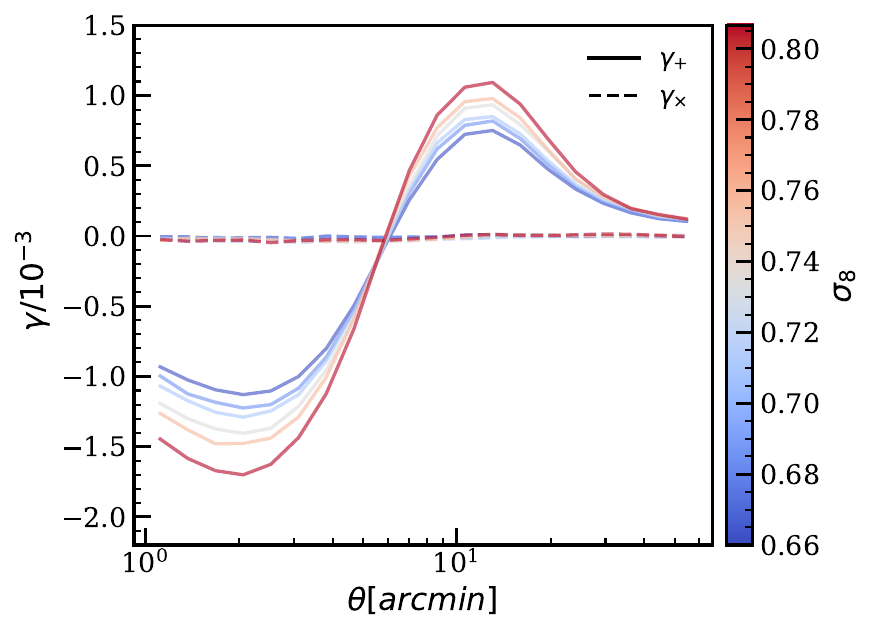}
    \end{subfigure}
    \hfill
    \begin{subfigure}{0.48\textwidth}
        \centering
        \includegraphics[width=\linewidth]{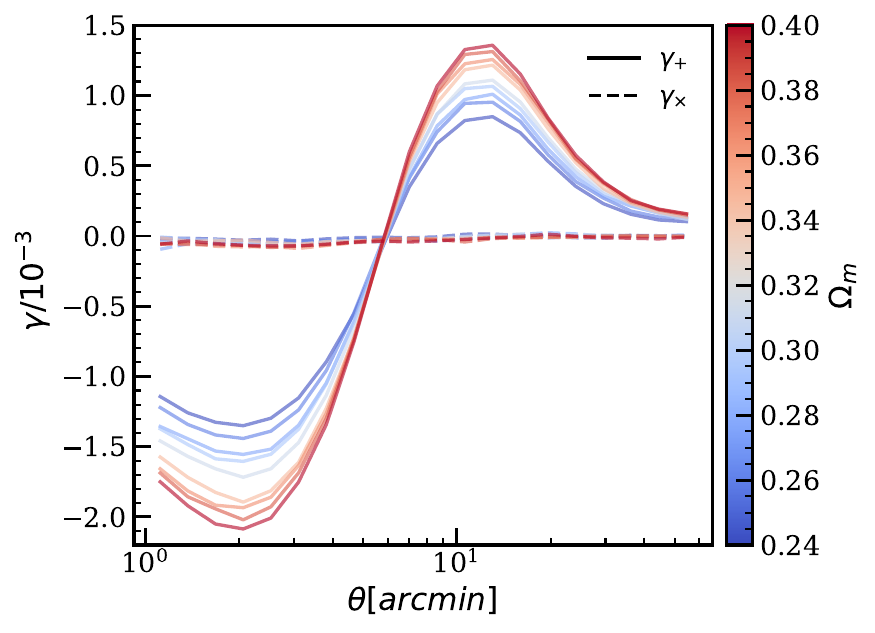}
    \end{subfigure}

    \caption{
        Behaviour of shear profiles $\gamma_{+}$ and $\gamma_{x}$ with respect to changing cosmological parameters.
        Top: $S_8$ and $S_8^{\perp}$.
        Bottom: fixed $\Omega_m$ and fixed $\sigma_8$.
    }
    \label{fig:shear_gplus_grid}
\end{figure*}

Figure \ref{fig:shear_gplus_grid} shows the dependence of the signal in the $\sigma_8$, $\Omega_m$, and $S_8$ directions, as well as the direction perpendicular to $S_8$ at the fiducial point.

A clear monotonic dependence of the shear signal amplitude on $S_8$ is observed, with higher values enhancing the signal in both the positive and negative regimes. There is no shift to the zero-crossing or other shape. In the perpendicular direction to $S_8$ there is only minimal variation of the signal at small scales, and no other clear effect. This change is within - as we shall see below - the noise levels for DES data. This suggests that ridge lensing may have similar general response to cosmic shear and therefore be of limited use as a higher-order statistic.  The two panels for $\Omega_m$ and fixed $\sigma_8$ show intermediate behaviour between the two, consistent with this interpretation.

\section{\label{sec:results2}Results: Noisy Simulations \&\\ Dark Energy Survey Data}

\begin{figure}[h!]
    \centering
    \includegraphics[width=0.45\textwidth]{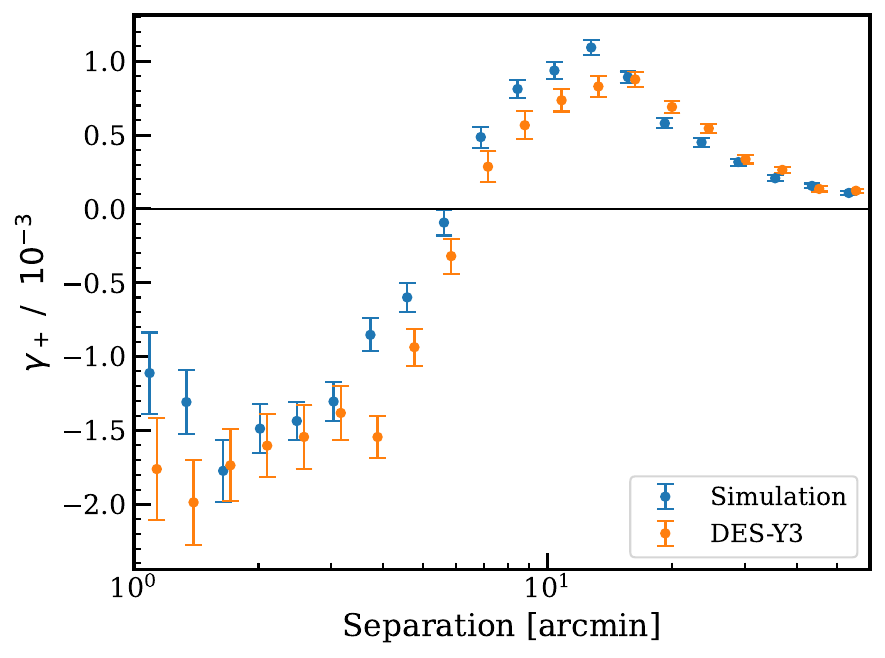}
    \caption{Ridge lensing tangential shear for Dark Energy Survey Year 3 data and a comparable noisy simulation with $\Omega_m=0.3$ and $\sigma_8=0.8$.}.
    \label{fig:desandsim}
\end{figure}

We now switch on the shape noise in the background shear simulation, with its level $\sigma_e$ matching the DES Y3 Metacalibration catalogs described in Section \ref{sec:data:des}, using our fiducial cosmological and other parameters. We also use the same configuration to run on real DES data (Metacalibration and Maglim), applying the $R$ calibration described in Section \ref{sec:data:des}. The resulting measurements are displayed in Figure \ref{fig:desandsim}.

\begin{figure}[h!]
    \centering
    \includegraphics[width=0.45\textwidth]{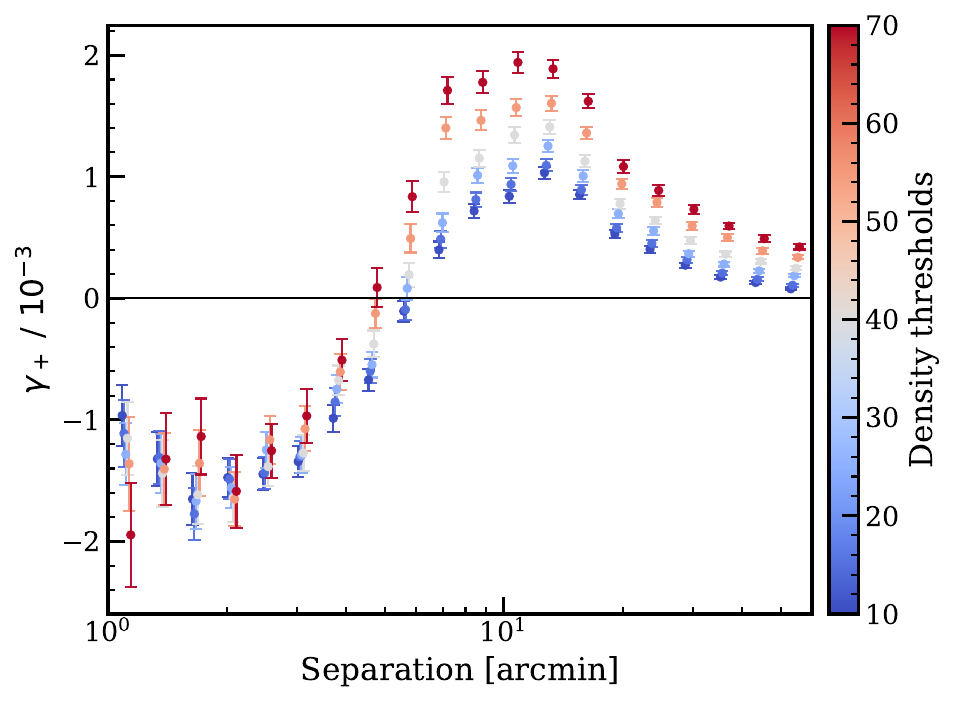}
    \caption{Ridge lensing tangential shear for increasing value of the threshold cut in noisy simulations}.
    \label{fig:noisyfp}
\end{figure}

The noise and signal are generally consistent in both simulation and data, and we are able to make a clear detection of the signal. They are also consistent, in both overall shape and amplitude, with the signal-only simulations in previous sections.

We assess the statistical significance of the fiducial simulation by computing the $\chi^2$ statistics for the stacked shear signal using an estimated covariance from realizations with randomly rotated shears. This yields a highly significant $\gamma _+ $ detection at a signal to noise ratio $S/N \approx 57$ while, the $\gamma _{\times}$ component remains consistent with zero.  The DES Y3 measurement yields a comparable tangential shear detection at $S/N \approx 47 $, and also has a cross component remains consistent with zero.  

While the choice of our fiducial values yield a clear detection in both simulation and the data the density threshold cut constitutes a non-trivial analysis choice. Increasing the value of the cut enhances the shear amplitude by preferentially selecting the higher density ridges but simultaneously reduces the number of contributing structures (lenses). The statistical significance should therefore have some maximum at an intermediate threshold. To quantify this, we have evaluated the $S/N$ of the tangential shear with respect to the increase in the density percentile shown in Figure ~\ref{fig:noisyfp}. This allows us to assess the robustness of the detection against variation in the density cut. The resulting $S/N$ shown in Table ~\ref{tab:snr_threshold} increases to reach a maximum at the 55th percentile before declining at a higher cut.

Finally, We have performed a test of algorithmic stability. Holding the catalogue and fiducial choices fixed, and varying only the random seed controlling the SCMS initialization, we obtain an ensemble of 50 shear profiles. The typical internal scatter produces fractional RMS $\approx 1\%$ with a maximum of $8.7\%$ in the lowest signal-to-noise bin. The internal covariance is therefore subdominant to statistical shape noise covariance and may be neglected.

\begin{table}[t]
\centering
\caption{Tangential shear detection significance as a function of density-threshold percentile.}
\label{tab:snr_threshold}
\begin{tabular}{cc}
\hline\hline
Density percentile & SNR$_+$ \\
\hline
10 & 52.72 \\
15 & 56.97 \\
25 & 64.23 \\
40 & 63.88 \\
55 & 80.49 \\
70 & 72.32 \\
\hline
\hline
\end{tabular}
\end{table}

\section{Conclusion} \label{sec:conclusion}

In this paper we have introduced a novel weak lensing approach: computing tangential shear measured around density ridges identified in photometric foreground samples. We used the Subspace-Constrained Mean Shift algorithm, followed by a segmentation pipeline and tangential shear measurement, and detected a clear signal in the Dark Energy Survey Year 3 data.  We have shown that the signal is robust under changes of convergence parameters such as mesh size and the density threshold. 

We have explored the expected behaviour of the underlying signal using \textsc{Glass} simulations, with respect to both cosmological parameters and algorithmic ones. We find that there is a clear dependence on structure formation parameters, primarily on the $S_8$ parameter, and have shown that we recover stronger signals when we restrict to higher density foreground ridges, as expected. We find that the bandwidth parameter in the algorithm has a strong effect on the structure of the signal.

While the signal has a distinct morphology from that of galaxy-galaxy lensing, there is significant overlap in their behaviour and thus in our detection significance should not be seen as completely independent of that phenomenon: the shear morphology converges towards the  galaxy-galaxy lensing signal at the limit of small bandwidth or large scale, In future work we will consider different approaches to subtraction of the galaxy-galaxy lensing signal to obtain a more pure detection.

While our simple simulations are sufficient to recover the general signal strength, to be able to use them for precision parameter estimation we will need to make several improvements. In particular, we must apply more realistic treatment of redshift cuts to improve the simulations, and incorporate lens sample weights into the ridge estimation to account for varying depth across the field. 

With such improvements, we can consider emulation or simulation-based inference approaches to using ridge lensing for cosmological parameter estimates.


\begin{acknowledgments}
We thank Shrihan Agarwal for sharing examples of using the \textsc{Glass} simulation code. We also wish to acknowledge the work undertaken by the Cosmostatistics Initiative at the 6$^\mathrm{th}$ COIN Residence Program (CRP\#6) in 2019, the modified DREDGE code of which provided a starting point for our extensions. 

For the purpose of open access, the author has applied a Creative Commons Attribution (CC BY) licence to any Author Accepted Manuscript version arising from this submission.

Author MN acknowledges support provided by UGrant-start under the financial task number 533‑BG60‑GS18‑25.

\end{acknowledgments}

\appendix

\section{Formal ridge definition}\label{app:ridge}

The subspace-constrained mean shift algorithm was first introduced by \citet{Ozertem_2011}, starting with an initial uniform layer of points in the desired parameter space to identify local principal curves. The algorithm has been investigated in more detail by \citet{Genovese_2014}, who define a ridge for a map $\xi : \mathbb{R}^d \rightarrow \mathbb{R}$ as
\begin{eqnarray}
    R = \{ x : ||G(x)|| = 0, \lambda_{d + 1} (x) < 0 \},
\end{eqnarray}
where $\lambda$ denote eigenvalues of the Hessian, and $x$ represents locations in which an omitted largest eigenvalue is positive and $G$ is zero. The latter is defined as\begin{eqnarray}
    G(x) = L(x) \nabla p(x) \ \mathrm{s.t.} \ L(x) \propto L(H(x)) = v' v'^{\top},
\end{eqnarray}
with $\nabla p(x)$ as the projected gradient of a probability function  $p : \mathbb{R}^d \rightarrow \mathbb{R}$. Here, $v'$ denotes the columns of $U(x)$ associated with the $d - 1$ smallest entries in $\lambda$ in an eigendecomposition with a diagonal matrix $\Lambda(x)$ featuring $\lambda$ as the diagonal,
\begin{eqnarray}
    H(x) = U(x) \Lambda(x) U(x)^{\top},
\end{eqnarray}
where $\lambda$ signifies eigenvalues sorted in descending order, corresponding to eigenvalues $v$. As the dimensionality of the space is $d$, the eigenvector running parallel to a given ridge is thus removed. A more detailed overview can be found in the foundational papers by \citet{Ozertem_2011} and \citet{Genovese_2014}.

\section{SCMS Algorithm Listing}\label{app:scms}
The SCMS algorithm steps are shown below. The pseudocode features the updated convergence criterion for individual mesh points described in Section~\ref{sec:scms_changes}.

\begin{algorithm}[H]
\caption{SCMS with thresholding, adapted from \citet{Moews_2020}.}
\begin{algorithmic}[1]
\State \textbf{Input:} Coordinates $\theta$, bandwidth $\beta$, threshold $\tau$, iterations $N$
\State \textbf{Output:} Density ridge point coordinates $\psi$
\Procedure {SCMS}{$\theta$, $\beta$, $\tau$, $N$, $T$}
\State $\kappa(x) \gets \mathrm{KDE}_{\mathrm{RBF}} (\theta, \beta)$
\State $x_{\mathrm{min}} \gets (\min(\theta_{\ast, 1}), \max(\theta_{\ast, 1}))$
\State $y_{\mathrm{min}} \gets (\min(\theta_{\ast, 2}), \max(\theta_{\ast, 2}))$
\State $\mathcal{\psi} \gets \mathcal{\psi} \sim U(x_{\mathrm{min}}, y_{\mathrm{min}})_{\mathrm{dim}(\theta)}$
\State $\psi \gets \forall y \in \psi : \kappa(y) > \tau$
\For {$i \gets 1, 2, \dots, \mathrm{dim}(\psi)$}
\For {$n \gets 1, 2, \dots, N$}
\For {$j \gets 1, 2, \dots, \mathrm{dim}(\theta)$}
\State $a_j \gets \frac{\psi_i - \theta_j}{\beta^2}$
\State $b_j \gets \mathcal{K} \left( \frac{\psi_i - \theta_j}{\beta} \right)$
\EndFor
\State $H(x) \gets \frac{1}{\mathrm{dim}(\theta)} \sum_{j = 1}^{\mathrm{dim}(\theta)} b_j \left( a_j a_j^{\top} - \frac{1}{\beta^2} \mathbb{I} \right)$
\State $v, \lambda \gets v, \lambda \mathrm{ \ from \ diagonalisation \ eig} (H(x))$
\State $v' \gets \mathrm{entries \ in \ } v \mathrm{ \ corresp. \ to \ sort}_{\mathrm{asc}} (\lambda)_{1, 2, \dots, d - 1}$
\State $\psi' \gets v' v'^{\top} \frac{\sum_{j = 1}^{\mathrm{dim}(\psi)} b_j \theta_j}{\sum_{j = 1}^{\mathrm{dim}(\psi)} b_j}$ 
\State $\Delta \gets \psi - \psi'$
\State $\psi_i \gets \psi'$
\If{$|\Delta| < T$}
break
\EndIf
\EndFor
\EndFor
\State \textbf{return} $\psi$
\EndProcedure
\label{alg:scms}
\end{algorithmic}
\end{algorithm}

\bibliography{main_shortened}

\end{document}